\documentstyle[12pt]{article}
\begin{document}
 \baselineskip = 24 true pt
\begin{center}
{\Large Interacting anyons and the Darwin Lagrangian}\\[1cm]
N. Banerjee$^1$ and Subir Ghosh$^2$\\[.5 cm]
$^1$ Saha Institute of Nuclear Physics, Sector 1, Block AF\\
Bidhannagar, Calcutta 700064, India.\\[.5cm]
$^2$ Physics Department, Gobardanga Hindu College,\\
Gobardanga, 24 Pgs.(North), West Bengal, India.
\end{center}
\vspace{1cm}
\begin{abstract}

We propose a new model for interacting (electrically charged) anyons, where the
2+1-dimensional  Darwin  term is responsible for interactions. The Hamiltonian
is comparable with the one used previously (in the RPA calculation).

\end{abstract}
\newpage
This letter aims at presenting a new model to describe an interacting anyon gas
\cite{W1,W2}. Although somewhat naive and intuitive, this model indeed produces
encouraging  results.  The  idea  is to start from a system of charged
particles,
interacting  via  the  conventional  Coulomb  forces,  in  2+1  dimensions.
The
free  part  of the Hamiltonian is simply a sum of free particle kinetic
energies.
For the assembly of such particles we incorporate the  interparticle
interaction
via  the  Darwin  term \cite{L}. Now we introduce the fact that the particles
are
actually anyons. For a free anyon, we use the spinning particle model proposed
by
one  of  us  \cite{G}.  The fundamental noncanonical commutator structure
derived
there \cite{G}, in the low energy limit,
is equivalent to that of Chou, Nair and  Polychronakos  \cite{CNP}.
We  use  a  noncanonical transformation, introduced in \cite{CNP} and analyse
the
resulting modified Hamiltonian, written in terms  of  canonical  variables.
This
Hamiltonian is able to reproduce not only the second quantised Hamiltonian of
the
RPA calculation \cite{W2}, but some correction terms as well.

It is amusing  to  note  that  in  some  sense,  the  present  formulation  is
complementary  to  that of \cite{W2}, so far as obtaining the Hamiltonian, the
starting point of RPA calculations \cite{W2} is concerned.
  In  the  latter  case,  one  starts from a Hamiltonian of particles
interacting with an external gauge field. However, the gauge  field  dynamics
is
governed  by  the  Chern-Simons  term  and  so the gauge potential (and hence
the
fields) are not independent and are constructed  from  the  particle
coordinates
\cite{J}.  After  using  this  known  form  of  gauge  potential and some
further
manipulations, one recovers the Coulomb interaction term in the second
quantised
Hamiltonian  of \cite{W2}. On the other hand here we start with a conventional
relativistic Darwin Lagrangian, construct the nonrelativistic (low energy)
Hamiltonian,
use  the  noncanonical transformations and finally end up with the Hamiltonian
of a system of charged  particles,  interacting  via  the  Chern-Simons  gauge
potential.

The spinning particle model \cite{G}  of  free  anyons,
and  the  present  interacting  system  as  well,  has  some  advantages  over
the
Chern-Simons constructions \cite{W1}. Originally the statistical gauge field,
that is the Chern-Simons field was introduced with the sole purpose of
modifying
the  spin-statistics  of the matter single particle states coupled to it.
However
in the relativistic context, as for example in our case, this mechanism does
not
work  \cite{JN}.  Also  as  pointed  out in \cite{JN}, even after eliminating
the
gauge  field  residual  interactions   might   still   persist.   Obviously
the
Chern-Simons construction is not the minimal way of generating anyons
\cite{CNP}.
{}From  a  technical  point of view, removal of the gauge field produces a
nonlocal
Lagrangian for a free anyon which is also  problematic.  All  these  problems
are
avoided in the spinning particle formulation \cite{G}, which is local and
consists of the
particle degrees of freedom only. The elegance of the Darwin construction is in
its simplicity. In this conceptually clear scheme, one straightaway derives the
relativistic
interacting Lagrangian. A precise approximation scheme is also available to
make
contact with existing results.

Let  us  now  formulate explicitly the 2+1-dimensional Darwin Lagrangian ($L$)
for a system of electrically charged spining particles. We will deal with the
two particle case, which obviously can be generalised to a many particle
system.
The
single particle  Lagrangian  $L_1$  for
particle(\#1) in presence of particle (\#2) is,
\begin{equation}
L_1 = L_1^{(0)} - e \phi_{12} + e \vec{A_{12}}.\vec{v_1}\label{1}
\end{equation}
where  $e$  is  the  particle  charge, $v_1$ is the velocity of particle 1 and
$\phi_{12}$   and  $\vec{A_{12}}$  are  the   scalar   and   vector
potentials
(retarded) due  to
particle  2, which is also moving,
on  particle  1. Note that we will consider a low energy approximation and the
small
 expansion parameters are $|\vec v|$ and $1/m$ with $p_0\approx m\gg |\vec v|$.
In our analysis we will retain terms upto $O(v^2)$ and $O(1/m^2)$.
 $L_1^{(0)}$ is the previously derived free anyon  Lagrangian  of
\cite{G}.

It  should be pointed out that the free  part of the Lagrangian,
$L_1^{(0)}$, will be treated in an identical fashion as in \cite{G},  in  the
subsequent
Hamiltonian  analysis.  We  will  compute Dirac brackets \cite{D} for the free
anyon part of the Hamiltonian only \cite{G} and then consider the  interaction
part,keeping in mind the non-canonical Dirac bracket structure of the
fundamental variables.
The 2+1-dimensional Maxwell equations are,
\begin{eqnarray}
\frac{\partial F_{\alpha\beta}}{\partial x^\beta} &=& -2\pi j^\alpha,\\
\frac{\partial\epsilon^{\alpha\beta\gamma}F_{\beta\gamma}}{\partial x^\alpha}
& =& 0,\\
F_{\alpha\beta} = \partial_{\alpha}A_\beta - \partial_\beta A_\alpha&,& j^\mu
= (\rho, \rho\vec v).
\end{eqnarray}
The retarded potentials are
\begin{eqnarray}
\phi_{12} &=& \int d^2x^2 \rho_2(t-R_{12})\ln R_{12} \nonumber\\
&=& \int d^2x^2\left[\rho_2 \ln R_{12} - \frac{\partial\rho_2}{\partial
t}R_{12}\ln
R_{12}  +  \frac{1}{2}\frac{\partial^2\rho_2}{\partial t^2} R^2_{12}\ln R_{12}
\right]\nonumber\\
&=& \int {d^2}{x^2}\rho_2\ln R -  \frac{\partial}{\partial  t}\int  \rho_2  R
\ln  R  +
\frac{1}{2} \frac{\partial^2}{\partial t^2}\int R^2\ln R,\label{2}\\
\vec{A}_{12}  &=&  \int d^2x^2 \rho_2\vec{v}_2\ln  R_{12}=  \int d^2x^2
\rho_2\vec{v}_2\ln
R.\label{3}
\end{eqnarray}
where  $\rho_2$ is the charge density due to second particle,  $\vec{R_{12}}
= \vec{x_1} - \vec{x_2}$ is the relative position of (1) with respect (2)
and we expanded $\rho_2$ around $t$. In the last lines of the
equations we put $R$ instead of $R_{12}$ and we use natural units.  Note  that
to  the  required  order, it is sufficient to retain the zeroth order term for
vector potential only. Assuming all the particles are pointlike, {\it i.e.}
$\rho_2 = \delta(\vec r - \vec x_2(t))$, the potentials are,
\begin{eqnarray}
\phi_{12} &=& e\ln R - e \frac{\partial}{\partial t}(R\ln R) + \frac{1}{2}e
\frac{\partial^2}{\partial t^2} (R^2\ln R)\label{4}\\
\vec A_{12} &=& e\vec v_2 \ln R\label{5}
\end{eqnarray}
For  convenience  we  shift  to  the  Coulomb  gauge  via  the following gauge
transformation \cite{L}
\begin{eqnarray}
\phi^\prime_{12}  &=&  \phi_{12}   -\frac{\partial   f}{\partial   t}~~,~~\vec
A^\prime_{12} = \vec A_{12} + \vec\nabla f,\label{6}\\
f &=& e(-R\ln R + \frac{1}{2} \frac{\partial}{\partial t}(R^2\ln R).\label{7}
\end{eqnarray}
The potentials are changed to
\begin{eqnarray}
\phi^\prime_{12} &=& e\ln R\label{8}\\
\vec A^\prime_{12} &=& e\vec v_2\ln R -e \hat n \ln R - e \hat n + \frac{e}{2}
\frac{\partial}{\partial t}(2R\ln R \hat n + R\hat n)\label{9}
\end{eqnarray}
$\phi^\prime$ has reduced to instantaneous Coulomb potential. However, $\vec A$
is more involved than its 3+1 dimensional counterpart \cite{L}. With the help
of
the relations, `dot' denoting time derivative,
\begin{equation}
\dot  R = -\frac{\vec R.\vec v_2}{R}~~,~~ \hat{\dot n}= \frac{-\vec v_2 + \hat
n (\hat n.\vec v_2)}{R},
\end{equation}
we finally have
\begin{eqnarray}
\vec A^\prime &=& -e\hat n (\ln R + 1 + \hat n.\vec v_2) -\frac{e}{2}\vec  v_2
\label{10}\\
\phi^\prime &=& e\ln R\label{11}
\end{eqnarray}
Let us straightaway move on to the two particle Hamiltonian $H$,
$$H  =  2(\frac{p^2}{2m}  - \frac{p^4}{8m^3}) + e^2 \ln R
+\frac{e^2}{2} [(\hat n_{ab}.
\frac{\vec p_a}{m})(\ln R +1+ \hat n_{ab}.\frac{\vec p_a}{m})$$
$$+ \frac{\vec p_a.\vec p_b}{2m^2}
+ (\hat n_{ba}.
\frac{\vec p_b}{m})(\ln R +1+ \hat n_{ba}.\frac{\vec p_b}{m})
+ \frac{\vec p_a.\vec p_b}{2m^2}]$$
\begin{equation}
  =  \frac{p^2}{m}  - \frac{p^4}{4m^3} + e^2 \ln R
+e^2[\frac{\vec  R.\vec  p}{mR}  (1+  \ln  R   +\frac{\vec   R.\vec   p}{mR})
-\frac{p^2}{2m^2}]\label{12}
  \end{equation}
Note  that  $\vec p  =  m\vec v$, and $\vec p_a = -\vec p_b = \vec p$  and
$\hat n_{ab}=
-\hat n_{ba} = \vec  R/R$.  We  also  reduced  the  relativistic  model   to  a
nonrelavistic  one  ($v\ll  1$)  and  $O(p^4)$ term is the standard correction
term, in the  limit  $p^0\simeq  m\gg  |\vec  p|$. But in our approximation
this term
of $O(mv^4)$ can be ignored.
As  mentioned  before  the
following  noncanonical  brackets  between  $p_a$  and  $x^b$  have already
been
calculated \cite{G}
\begin{eqnarray}
\{ p^\mu, x^i\} &=& -g^{\mu i} + \frac{g^{\mu0}p^i}{m}\label{13}\\
\{p^\mu, p^\nu\} &=& 0\label{14}\\
\{x^i,x^j\}  &=&  -\frac{2J}{m^2}\epsilon^{ij}  +
\frac{2J}{m^4}(\epsilon^{jk}p^i -
\epsilon^{ik}p^j)p_k\label{15}
\end{eqnarray}
These Dirac brackets include the gauge fixation of $x^0$ to be the time.  This
makes  the  $x^0$ brackets trivial. Also we have written the expression in the
low energy domain by putting $p_0\simeq m$. Finally note that in  the  result
in  \cite{CNP}  only  the  $O(\frac{1}{m})$  term is in $\{x^i,x^j\}$-bracket
is
present. As mentioned before we can
ignore the last term in (\ref{15}), which is of $O(\frac{v^2}{m^2})$ in
subsequent
analysis.

Let  us  now  introduce  the  non-canonical  transformations,  first  used  in
\cite{CNP}. Actually  these  are  nothing  but  solutions  of $\vec p$ and
$\vec x$,
in terms of a canonically conjugate pair $\vec p$ and $\vec q$ which are
consistent with the  Dirac
brackets.  We  will write the Hamiltonian in terms of the new set of canonical
variables. The above mentioned transformations are,
\begin{eqnarray}
x^i &=& q^i + \frac{J\epsilon^{ij}p_j}{m^2}\label{16}\\
p_i&=& p_i\label{17}
\end{eqnarray}
It seems that the set of relations are analogus to those used in \cite{J}, the
difference being that  in  \cite{J},  one  modifies  the  momenta  keeping  the
coordinate unchanged, whereas it is the reverse in the present case. Thus we
have,
\begin{eqnarray}
R^i &=& q^i +\frac{2J}{m^2}\epsilon^{ij}p_j\label{17a}\\
\ln R &=& \ln q + \frac{2J}{m^2q^2}\epsilon^{ij}q_ip_j\label{17b}\\
(\frac{1}{R^2})^{1/2} &=& \frac{1}{|q|}(1-
\frac{2J}{m^2q^2}\epsilon^{ij}p_jq_i)\label{17c}\\
\vec R.\vec p &=& \vec q.\vec p
\end{eqnarray}
It is now straightforward to express the Hamiltonian in terms of $\vec q$  and
$\vec p$
and we obtain
\begin{equation}
H  =  \frac{1}{m}  (\vec  p  - e\vec a)^2 +  e^2 [\ln q + \frac{\vec q.\vec
p}{mq}(1+\ln q) ] +\frac{e^2}{m^2}[(\frac{\vec q.\vec p}{q})^2-  \frac{p^2}{2}]
-\frac{e^2{\vec
a}^2}{m}\label{18}\end{equation}
where
\begin{equation}
a^i  =  \frac{Je}{m}\frac{\epsilon^{ij}q_j}{q^2}  (1-  \frac{\vec  q.\vec
p}{qm}
\ln q)\label{19}
\end{equation}
The last term, being of $O(e^2/m^3)$ can be safely ignored.  One
can  at  once  see the striking result that this $a^i$ is nothing but the gauge
potential, that
is obtained from solving the theory of a  free  particle  interacting  with  a
Chern-Simons  gauge  field.  The  Chern-Simons  coupling  $k$ is identified as
$\frac{1}{2\pi k} = \frac{Je}{m}$. However  there  is  a  correction  term  to
$a^i$.  The  other  interesting  feature  of  our  scheme is the presence of
the
Coulomb  interaction  terms.  This  Hamiltonian  is  equivalent  to  the
Hamiltonian  from which the RPA calculations start \cite{W2},  although
obtained
from  a  completely different
starting point, the Darwin Hamiltonian.

It would be very nice to estimate the effect of the correction to $\vec a$ in
some physically relevant parameter. Note that this term is capable of changing
the sign of the gauge interaction in the Hamiltonian.

An interesting open problem is the construction of the  Lagrangian  that  will
directly  reproduce  the  final Hamiltonian(\ref{18}). This is not the same as
the Darwin Lagrangian because the noncanonical transformations have mixed  the
coordinates  and  momenta.  This  Lagrangian will be useful in formulating the
problem of a relativistic anyon gas, in a covariant framework.

Another  problem  worth  looking   at  is  the  Dirac  analysis  of  the  full
interacting  system.  (Remember that here we have done the constraint analysis
of the {\it free} spinning particle system.) Also in this  respect  one  might
try  to compute the dipole-dipole interaction correction, originating from the
parent spinning charged particle.

\end{document}